\documentclass[10pt,conference]{IEEEtran}
\IEEEoverridecommandlockouts
\usepackage{amsmath,amssymb,amsfonts}
\usepackage{algorithmic}
\usepackage{graphicx}
\usepackage{textcomp}
\usepackage{xcolor}
\usepackage{nicefrac}
\usepackage{booktabs}
\usepackage[hidelinks]{hyperref}
\usepackage[sort, numbers]{natbib}
\usepackage{inconsolata}
\def\BibTeX{{\rm B\kern-.05em{\sc i\kern-.025em b}\kern-.08em
    T\kern-.1667em\lower.7ex\hbox{E}\kern-.125emX}}

\begin{document}
\begin{sloppy}

\title{She Elicits Requirements and He Tests: Software Engineering Gender Bias in Large Language Models}

\author{\IEEEauthorblockN{Christoph Treude}
\IEEEauthorblockA{The University of Melbourne, Australia\\
christoph.treude@unimelb.edu.au}
\and
\IEEEauthorblockN{Hideaki Hata}
\IEEEauthorblockA{Shinshu University, Japan\\
hata@shinshu-u.ac.jp}
}

\maketitle

\begin{abstract}

Implicit gender bias in software development is a well-documented issue, such as the association of technical roles with men. To address this bias, it is important to understand it in more detail. This study uses data mining techniques to investigate the extent to which 56 tasks related to software development, such as assigning GitHub issues and testing, are affected by implicit gender bias embedded in large language models. We systematically translated each task from English into a genderless language and back, and investigated the pronouns associated with each task. Based on translating each task 100 times in different permutations, we identify a significant disparity in the gendered pronoun associations with different tasks. Specifically, requirements elicitation was associated with the pronoun ``he'' in only 6\% of cases, while testing was associated with ``he'' in 100\% of cases. Additionally, tasks related to helping others had a 91\% association with ``he'' while the same association for tasks related to asking coworkers was only 52\%. These findings reveal a clear pattern of gender bias related to software development tasks and have important implications for addressing this issue both in the training of large language models and in broader society.

\end{abstract}

\section{Introduction}

Implicit gender bias is prevalent among professional software developers. For instance, a study of 142 professional software engineers undertaking an Implicit Association Test~\cite{wang2019implicit} found significant associations between men and technical leadership positions, general technical positions, and career advancement. Garcia et al.~\cite{garcia2022gender} discovered that \textit{pink tasks}, which are tasks that require high standards and timeliness but offer little substantive development or visibility~\cite{brough2011women}, were often assigned to female students in team projects. Terrell et al.~\cite{terrell2017gender} found that while pull requests submitted by women tend to be accepted more often than those submitted by men, this is only the case when the women's identities are not immediately apparent.

Gender bias can lead to a lack of representation and opportunities for underrepresented groups, which can negatively impact innovation and productivity. For example, Vasilescu et al.~\cite{vasilescu2015gender} found gender diversity to be a significant positive predictor of productivity in GitHub teams. Gender bias can also perpetuate discrimination and create a hostile work environment, leading to high turnover rates with the associated knowledge loss~\cite{robillard2021turnover} and a lack of diversity in the workforce. 

To effectively address these biases, it is crucial to understand them in more detail. By identifying specific tasks and activities that are affected by gender bias, we can target our efforts to eliminate bias more effectively. Understanding the nuances and complexities of bias, such as how it can manifest differently depending on the context, is crucial to creating an inclusive and equitable software development community.

The advent of large language models has given us a new powerful tool to study such gender bias based on data mining techniques rather than expensive surveys and experiments. Large language models process vast amounts of natural language data, making it possible to identify patterns of gendered language that may not be immediately apparent when looking at raw data, such as detecting problematic associations using Caliskan et al.'s Word Embedding Association Test~\cite{caliskan2017semantics} or investigating the use of gendered pronouns in association with certain tasks or positions~\cite{bordia2019identifying}. 

In this study, we use back-translation~\cite{prabhumoye2018style} to mine gender bias associated with 56 tasks related to software development, taken from previous work~\cite{masood2022like}. We use the pronoun ``she'' to describe each task in English, translate the tasks into the genderless language of Finnish, and back into English using the DeepL translator.\footnote{\url{https://www.deepl.com/translator}} We then analyse the pronouns in the result. To account for the impact of context on translation, we perform 100 permutations of the task list and aggregate the results.

We identify a significant disparity in the gendered pronoun associations with different tasks. For instance, the sentence ``As a software engineer, she elicits requirements'' was changed to ``As a software engineer, he elicits requirements'' in only 6\% of all cases after round-trip translation (with ``he or she'' and ``he/she'' accounting for the remaining cases), while ``As a software engineer, she tests'' was changed to ``As a software engineer, he tests'' in 100\% of the cases. Other tasks with a relatively weak association with ``he'' included task estimation, infrastructure setup, and support tasks, while providing comments and learning were exclusively associated with ``he''.

Our findings reveal the specific software development tasks and activities that are most affected by gender bias, which can help everyone involved in software projects anticipate and proactively address potential issues. For example, role assignments should be made with conscious effort to avoid reinforcement of stereotypes. Additionally, mining techniques can be used to uncover patterns of gendered language in software repositories and internal communication, and necessary adjustments can be made to eliminate bias. Our work also highlights the importance of addressing bias in large language models at a fine-grained task level, in addition to efforts aimed at addressing more coarse-grained biases related to social~\cite{liang2021towards}, political~\cite{liu2021mitigating}, and sentiment~\cite{huang2020reducing} aspects.

\section{Related Work}

The biases that systems can contain have been studied in areas such as machine learning, natural language processing, and deep learning.
\citet{caliskan2017semantics} showed for the first time that applying standard machine learning to ordinary language results in human-like semantic bias.
For biases in text corpora, \citet{bordia2019identifying} proposed a metric to measure gender bias and a method to reduce gender bias.
\citet{10.1145/3457607} investigated real-world applications that demonstrated bias in different ways and listed different sources of bias that could affect artificial intelligence applications. They also examined different domains and sub-domains in AI and showed how researchers have observed and tried to address unfair results using state-of-the-art methods.

Various empirical studies have been conducted on gender bias in software development.
\citet{6542459} assessed gender representation and social impact on Stack Overflow and reported that the majority of contributors to Stack Overflow are men, and that men have gained more reputation.
\citet{10.1007/978-3-642-33442-9_6} examined the mailing list subscription and posting statistics of FOSS participants in six FOSS projects and found that the participation rate of women is decreasing.
Comparing the acceptance rates of contributions from men and women in the open source software community, \citet{terrell2017gender} reported that women's contributions tend to be accepted more frequently than men's, while for contributors who are outsiders to the project and whose gender is identifiable, the acceptance rate is higher for men.
From an analysis of the impact of gender bias in the GitHub, \citet{10.1109/ICSE.2019.00079} reported that women provide less information, work on fewer projects and organisations, and are more restrictive in expressing their sentiments.
\citet{9466393} conducted a multi-regional geographic analysis of gender inclusion in GitHub and reported that gender diversity is low in all regions of the world, with no substantial differences by region.
In an empirical study of newcomer women, \citet{9055190} reported that female newcomers encountered gender bias in 63\% of the barriers they faced.
\citet{10.1145/3510458.3513011} analyzed the development histories of a large number of software projects and reported that while the percentage of female authors worldwide is low but steadily increasing, the percentage of female participation declined during the COVID-19 epidemic.

This study is unique in its subject and approach: it analyzes fine-grained tasks in software development from the perspective of current large language models.

\begin{table*}
\centering
\caption{Results indicating the frequency of pronoun translation for each sentence, broken down by ``she'', ``he/she'', ``he or she'', and ``he''. The percentage of instances in which the pronoun was translated as ``he'' is provided in the final column. Note that a small number of translations did not produce any pronoun, and that all original sentences began with ``As a software engineer, ''.}
\label{tab:results}
\begin{tabular}{l|rrrr|r}
\toprule
\textbf{Original Sentence} & \textbf{``she''} & \textbf{``he/she''} & \textbf{``he or she''} & \textbf{``he''} & \% \textbf{``he''}\\
\midrule
She elicits requirements. & 0 & 51 & 43 & 6 & 6\% \\
She estimates tasks/projects. & 0 & 61 & 0 & 39 & 39\% \\
She performs infrastructure setup. & 0 & 39 & 14 & 47 & 47\% \\
She performs support tasks. & 0 & 44 & 6 & 49 & 49\% \\
She archives code versions. & 0 & 16 & 34 & 50 & 50\% \\
She generates reports/documents. & 0 & 47 & 0 & 51 & 52\% \\
She submits changes. & 0 & 23 & 26 & 51 & 51\% \\
She asks coworkers. & 0 & 46 & 2 & 52 & 52\% \\
She performs administrative tasks. & 0 & 35 & 7 & 57 & 58\% \\
She performs personal debugging. & 0 & 38 & 5 & 57 & 57\% \\
She performs user training. & 0 & 28 & 9 & 63 & 63\% \\
She stores design versions. & 1 & 11 & 21 & 67 & 67\% \\
She assigns GitHub issues. & 2 & 22 & 6 & 69 & 70\% \\
She manages development branches. & 1 & 17 & 8 & 74 & 74\% \\
She mentors others. & 0 & 13 & 13 & 74 & 74\% \\
She browses FAQs. & 0 & 24 & 0 & 76 & 76\% \\
She browses documentation. & 0 & 19 & 3 & 77 & 78\% \\
She commits code. & 0 & 14 & 7 & 78 & 79\% \\
She reviews pull requests. & 0 & 14 & 6 & 80 & 80\% \\
She assesses potential problems. & 0 & 18 & 1 & 81 & 81\% \\
She fixes bugs. & 0 & 16 & 2 & 82 & 82\% \\
She reads/reviews code. & 0 & 18 & 0 & 82 & 82\% \\
She has meetings. & 0 & 16 & 1 & 83 & 83\% \\
She navigates code. & 0 & 14 & 3 & 83 & 83\% \\
She reads changes. & 0 & 17 & 0 & 83 & 83\% \\
She edits code. & 0 & 13 & 3 & 84 & 84\% \\
She edits artifacts. & 0 & 12 & 3 & 85 & 85\% \\
She writes documentation/wiki pages. & 0 & 14 & 0 & 86 & 86\% \\
She accepts changes. & 0 & 10 & 2 & 87 & 88\% \\
She produces on-line help. & 0 & 9 & 2 & 87 & 89\% \\
She submits pull requests. & 0 & 11 & 2 & 87 & 87\% \\
She classifies requirements. & 0 & 12 & 0 & 88 & 88\% \\
She inspects code. & 0 & 9 & 2 & 88 & 89\% \\
She networks. & 0 & 11 & 1 & 88 & 88\% \\
She provides comments on project milestones. & 0 & 11 & 0 & 89 & 89\% \\
She fixes defects. & 0 & 8 & 2 & 90 & 90\% \\
She provides comments on commits. & 0 & 10 & 0 & 90 & 90\% \\
She helps others. & 0 & 9 & 0 & 91 & 91\% \\
She produces user documentation. & 0 & 0 & 1 & 91 & 99\% \\
She provides enhancements. & 0 & 8 & 0 & 92 & 92\% \\
She releases code versions. & 0 & 7 & 1 & 92 & 92\% \\
She browses the web. & 0 & 6 & 0 & 94 & 94\% \\
She maintains changes. & 0 & 2 & 0 & 94 & 98\% \\
She reads artifacts. & 0 & 5 & 0 & 95 & 95\% \\
She produces tutorials. & 0 & 1 & 0 & 96 & 99\% \\
She browses articles. & 0 & 3 & 0 & 97 & 97\% \\
She identifies constraints. & 0 & 1 & 0 & 97 & 99\% \\
She codes. & 0 & 2 & 0 & 98 & 98\% \\
She plans. & 0 & 2 & 0 & 98 & 98\% \\
She writes emails. & 0 & 2 & 0 & 98 & 98\% \\
She removes dead code. & 0 & 1 & 0 & 99 & 99\% \\
She restructures code. & 0 & 1 & 0 & 99 & 99\% \\
She writes artifacts. & 0 & 1 & 0 & 99 & 99\% \\
She learns. & 0 & 0 & 0 & 100 & 100\% \\
She provides comments on issues. & 0 & 0 & 0 & 100 & 100\% \\
She tests. & 0 & 0 & 0 & 100 & 100\% \\
\bottomrule
\end{tabular}
\end{table*}

\section{Research Method}

In this section, we outline our research question and describe our methods for data collection and data analysis.

\subsection{Research Question}

Our research question is:

\begin{description}
\item[\textbf{RQ}] What implicit gender biases are embedded in large language models about software engineering tasks?
\end{description}

We aim to study software engineering gender bias in large language models and identify software engineering tasks that are typically associated with a specific gender.

\subsection{Data Collection}

To answer our research question, we needed a list of the various activities that fit into a software developer's job description. The most comprehensive list we identified was recently published by Masood et al.~\cite{masood2022like}. As part of their work, the authors surveyed literature on software development tasks and activities, and synthesized a list of example tasks and their categorization, primarily building on the work of \citet{meyer2019today}, \citet{licorish2017exploring}, \citet{graziotin2017unhappiness}, \citet{ford2015exploring}, \citet{milewski2007global}, \citet{glass1992software}, \citet{murgia2014developers}, and \citet{madampe2020towards}. We manually analysed the examples (from Table I of Masood et al.~\cite{masood2022like}) and split them into separate activities where applicable, e.g., extracting the activities ``assign GitHub issues'' and ``review pull requests'' from the example ``Assigning GitHub issue or reviewing pull request'', for a total of 56 software engineering tasks.

To identify bias, we relied on back-translation with a genderless language, i.e., a language that does not have grammatical gender, meaning that nouns, pronouns, and adjectives are not assigned a gender. We chose Finnish as the genderless language~\cite{renstrom2022gender}. As an example, the translation of the phrases ``she is a software engineer'' and ``he is a software engineer'' both results in the Finnish phrase ``hän on ohjelmistoinsinööri'' since the Finnish word ``hän'' is the translation of both ``she'' and ``he''. When translating the Finnish phrase back into English, the translator needs to decide whether to translate ``hän'' into ``she'' or ``he'', and translators that rely on large language models tend to use their implicit gender biases to make this decision.

We chose the DeepL translator for our study. DeepL advertises itself as ``the most accurate and nuanced machine translation''\footnote{\url{https://www.deepl.com/en/whydeepl}} and provides a free trial and an API. We were not able to use alternatives such as Google Translate or ChatGPT for our study since these tools have built-in heuristics to avoid biased output when translating from a genderless language to a gendered one, such as providing both translations (Google Translate) or providing disclaimers (ChatGPT).

Through our first experiments with DeepL, we noticed that the gendered pronoun in the final result depends on the surrounding sentences. For example, if only a single task was given, ``she'' was always back-translated into ``he''. However, if multiple tasks were given in the same input, some tasks were assigned to ``she'', some were assigned to ``he/she'', some were assigned to ``he or she'', some were assigned to ``he'', and for a small minority, all pronouns were omitted. To systematically study this, we provided DeepL with the list of 56 tasks 100 times, randomly shuffling the order of tasks each time.

Several tasks, such as ``accept changes'', would have a different meaning outside of a software engineering context. Therefore, we needed to ensure that the software engineering context was injected into the translation process. We experimented with different prefixes and found that the short phrase ``As a Software Engineer'' was not changed during back translation; it still said ``As a Software Engineer'' after translating in both directions. Since ``As a Software Engineer'' is not a task, we believe that this prefix would not introduce task-related bias. We also used ``she'' as the initial pronoun in all tasks, e.g., for the task ``help others'', the complete phrase given to DeepL was ``As a software engineer, she helps others.'' This provided us with a total of 5,600 translation results, i.e., 100 back-translations for each of the 56 phrases.

\subsection{Data Analysis}

To analyse the data, we aggregated for each of the 56 input sentences how often it had been back-translated using each pronoun variant (``she'', ``he/she'', ``he or she'', ``he'') across the 100 runs. Since ``he'' was by far the most common pronoun in the back-translations (in $\nicefrac{4,490}{5,600}=80.2\%$ of all sentences), we then calculate for each input sentence how often it was back-translated using the pronoun ``he'', compared to all other translations.

\subsection{Data Availability}

We make our data and scripts available in our online appendix:\footnote{\url{https://doi.org/10.5281/zenodo.7745436}} \texttt{tasks.txt} contains all sentences used as input for the back-translation, \texttt{llm-bias.py} contains the Python code that uses the DeepL API to back-translate each sentence 100 times, \texttt{output.txt} shows the raw output, and \texttt{llm-bias.csv} aggregates the results.

\section{Results}

Table~\ref{tab:results} shows the results. While the majority of tasks were translated using the pronoun ``he'', several tasks were often translated using ``he/she'' or ``he or she'', with only four out of 5,600 translations using the ``she'' pronoun that was originally given to the translator. The results indicate a significant disparity in the gendered pronoun associations with different tasks. Four tasks (elicit requirements, estimate tasks/projects, perform infrastructure setup, perform support tasks) were associated with ``he'' in the minority of cases, with ``he/she'' being prominent. On the other end of the spectrum, six tasks were associated with ``he'' in at least 99 out of 100 runs (remove dead code, restructure code, write artifacts, learn, provide comments on issues, test).

If we consider Masood et al.'s grouping of tasks into 14 sub-categories~\cite{masood2022like}, it is interesting to note that tasks related to requirements (elicit requirements 6\%, estimate tasks/projects 39\%, classify requirements 88\%) have a weaker association with ``he'' on average than any other grouping of tasks, followed by tasks related to documents and their maintenance (generate reports/documents 52\%, store design versions 72\%, maintain changes 98\%). On the other hand, tasks related to providing comments are more strongly associated with ``he'' on average than any other grouping (provide comments on issues 100\%, provide comments on commits 90\%, provide comments on project milestones 89\%).

For tasks that are directly related to each other, we observe several biases: asking coworkers has a relatively weak association with ``he'' (52\%) while helping others has a much stronger association (91\%). Browsing documentation and FAQs has a weaker association with ``he'' (78\% and 76\%, respectively) than browsing articles and the web (97\% and 94\%, respectively). Accepting changes has a stronger association with ``he'' (88\%) than submitting them (51\%). For artifacts, the strength of the association with ``he'' decreases from writing (99\%) and reading (95\%) to editing (85\%).

\section{Discussion}

Our analysis reveals a significant disparity in the gendered pronoun associations with various software development tasks. Our findings indicate that the majority of tasks are translated using the pronoun ``he'' more often than ``she''. However, the extent to which the large language model associates tasks with ``he/she'' or ``he or she'' instead of ``he'' varies. By being aware of these tendencies, individuals involved in software projects can take proactive steps to prevent and address potential issues. This includes being mindful of role assignments and making conscious efforts to avoid reinforcing stereotypes.

Addressing bias in language is a complex and ongoing process that requires a multifaceted approach. While some tools, such as Google Translate and ChatGPT, have implemented heuristics to suppress biased output in extreme cases, those do not address the underlying problem of bias in training data. Additionally, heuristics are unlikely to be complete since different grammatical forms could still expose biases. 

Data mining techniques can be used to extract implicit biases from large language models to better understand what needs to be mitigated, such as by providing models with large amounts of unbiased text data and fine-tuning them to recognize and correct for gender bias. Our research highlights that even within the single domain of software engineering, different but related tasks (e.g., ``elicit requirements'' and ``classify requirements'') have varying associations. 

While we cannot say for certain what caused the differences reported here, we can speculate based on related work and the other results we found. For example, the difference between ``elicit requirements'' (6\% ``he'') and ``classify requirements'' (88\% ``he'') might be caused by the fact that the former may involve talking to customers while the latter can be done without customer involvement. The association of ``she'' with customer-facing tasks is consistent with related work indicating that ``she'' is more likely to be associated with roles that require communication~\cite{fishman1997interaction}. This finding is also in line with our other results, which show that tasks that require communication are more closely related to ``she'' (e.g., mentor others, have meetings). More work is needed to mitigate such biases in large language models.

\section{Threats to Validity}

The validity of our results is affected by several threats. Our findings are specific to back-translations with one specific genderless language and the large language model used by the DeepL translator. It is important to note that all large language models may contain biases, despite some user interfaces attempting to mask such biases. Other ways of identifying biases, such as using a different genderless language, might have revealed different results.

The identified biases are dependent on the exact phrasing of each task. To minimize the introduction of additional subjectivity, we reused the phrasing from related work~\cite{masood2022like} as much as possible. However, we observed that the exact translation of each sentence was influenced by the surrounding sentences, and while we attempted to mitigate this by running the translation 100 times in different permutations, we cannot guarantee that 100 runs were sufficient. The number 100 was chosen based on the API limits of the DeepL translator.

We did not delve into the role of the combined pronouns ``he/she'' and ``he or she'' which accounted for 19.3\% of the output. Future research should investigate the intricacies of large language models that return combined pronouns and the reasons for the association of these forms with certain tasks. The same applies to other pronouns, such as ``his'', ``her'', and ``they'', among others. We plan to investigate these in future work.

\section{Conclusion and Future Work}

In this work, we used data mining techniques to examine the prevalence of implicit gender bias in software development tasks as manifested in large language models. Our analysis revealed a significant disparity in the gendered pronoun associations with various tasks, with a distinct pattern of bias observed in certain tasks, such as requirements elicitation. These findings have important implications for addressing gender bias in the training of large language models and in broader society, highlighting the importance of understanding and addressing this issue in more detail.

We will extend this work by considering other genderless languages, other translators, other pronouns, and other ways of phrasing software development tasks. In addition, we aim to conduct a detailed analysis of where these biases originate, such as project-specific documents or general-domain literature. We also plan to conduct a deeper analysis of the impact of the domain ``software engineering'' on these biases, focusing specifically on tasks that also exist in other domains, such as mentoring and planning.

\bibliographystyle{IEEEtranN}
\bibliography{msr23-llm-bias}

\end{sloppy}
\end{document}